\documentstyle[epsf,psfig,prl,twocolumn,aps]{revtex}

\begin{document}

\draft
\input epsf
\twocolumn[\hsize\textwidth\columnwidth\hsize\csname
@twocolumnfalse\endcsname

\title{Skyrme crystal or Skyrme liquid?}
\author{B. Paredes and J. J. Palacios}
\address{Dept. de F\'{\i}sica Te\'orica de la Materia Condensada,
Universidad Aut\'onoma de Madrid, Cantoblanco, Madrid 28049, Spain}

\date{\today}

\maketitle

\widetext
\begin{abstract}
\leftskip 2cm
\rightskip 2cm

We address the quantum melting phase transition
of the Skyrme crystal. Based on generic sum rules 
for two-dimensional, isotropic electron quantum liquids in the lowest 
Landau level, we propose analytic expressions for the pair distribution 
functions of spin-polarized and spin-unpolarized liquid phases at
filling factors $2/3\leq\nu\leq 1$. From the pair distribution functions we
calculate the energy of such liquid phases and compare with  
the energy of the solid phase. The comparison suggests that the quantum 
melting phase transition may lie much closer to $\nu=1$ than ever expected.

\end{abstract}

\vskip2pc]

\narrowtext

Whether or not a solid phase composed of skyrmions\cite{Sondhi}
exists near filling factor $\nu=1$ poses, probably, one of the most 
fascinating problems in
recent history of the quantum Hall effect\cite{Chakrabortybook}. This
phase, known as Skyrme crystal\cite{Brey}, is, nowadays, the target of a
number of experiments based on various techniques which, in the past, 
have borne out the existence 
of the skyrmions themselves\cite{Barrett,jpe-skyrme,Bayot}.
To date, however, little experimental evidence of the existence of the 
Skyrme crystal has been provided. Bayot et al.\cite{Bayot} have
reported a peak in the specific heat at very low temperatures which may
be attributed to a phase transition from a finite-temperature liquid  phase
into a solid phase at zero temperature\cite{Bayot,crystals}. 
The spin polarization measured by Barret et
al.\cite{Barrett} seems to be consistent, at least in the range of filling 
factors $|1-\nu|\approx 0.2$, with that calculated by Brey
et al.\cite{Brey} for a proposed solid phase. This apparent 
consistency neither guarantees that a zero-temperature 
liquid phase exists out of that range 
nor, most importantly, rules out its existence close to $\nu=1$; 
for an alternative proposal of an unpolarized or partially
polarized quantum liquid is not available to compare with. 

If a solid phase really exists at zero temperature, the following question
needs to be answered: What is the filling factor at
which quantum fluctuations turn the solid into a liquid?. Filling
factors $\nu=2/3$ and $\nu=4/3$ set the lower and upper 
melting transition limits, respectively.
(We will focus on $\nu < 1$ since the physics for $\nu > 1$ 
is similar due to particle-hole symmetry.) For large enough Zeeman 
energy per electron, $g\mu_BB/2$, 
the $\nu=2/3$ state is a spin-polarized liquid. More precisely, it is 
a $\nu=1/3$ Laughlin state of holes. For $g\mu_BB/2=0$ there has been mounting
numerical and experimental 
evidence of a spin-unpolarized liquid ground state\cite{Maksym}. 
To zero in on the actual location of the melting transition 
one can try to estimate the energy scale of the 
quantum fluctuations associated to the phonon modes in the classical 
solid phase\cite{crystals} and compare it with some relevant energy
scale, typically, set by the Coulomb interaction. 
This is a very crude approximation; moreover, it does
not provide any information about the liquid phase. Another approach,
used in Wigner crystal studies\cite{Yi} 
and which we explore in this Letter, is to offer an alternative
description of the ground state 
in terms of a quantum liquid, obtain its energy,
and compare it with that of the solid phase.

\begin{figure}
\vspace{-0.3cm}
\centerline{\epsfysize=7cm \epsfxsize=8.5cm \epsfbox{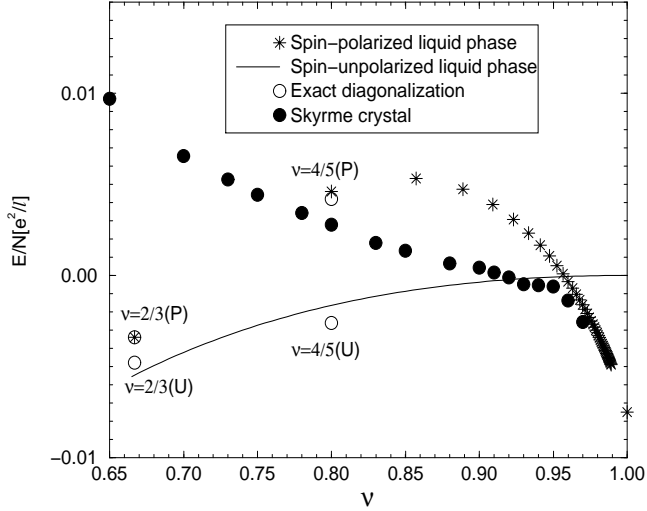}}
\caption{Energy per particle in the Hartree-Fock approximation
of the solid phase or Skyrme crystal (dots) 
and of the polarized (stars) and unpolarized
(solid line) liquid phases proposed in the text. 
For completeness, the values obtained from the exact diagonalizations are
presented by circles for both polarized and unpolarized phases. 
The results for the polarized liquid phase include a  
Zeeman energy shift of $g\mu_BB/2=-0.0075 e^2/\ell$ and all the energies 
are referred to $-\frac{1+\nu}{4}\sqrt{\pi/2}$.} 
\label{liquid}
\end{figure}

After Laughlin's proposal of a wave function for the main filling factors
$\nu=1/M$ (where $M$ is an odd integer $\geq 3$), many other wave
functions have been proposed for other filling factors with notorious
success\cite{Chakrabortybook}. For many purposes, however, a many-body wave 
function contains redundant information which can be compacted in a more 
advantageous way. In this spirit we note  that 
the ground state energy per particle, $E/N\equiv\epsilon$, for an infinite, 
neutral, and isotropic liquid state 
can be obtained from the pair distribution function 
$g(r)$\cite{Chakrabortybook}:
\begin{equation}
\epsilon=\frac{\nu}{2}\int_0^\infty \; dr \; r [g(r)-1] V(r),
\end{equation}
where $V(r)$ is the interaction potential. In general, 
$g(r)$ can only be obtained from the wave function using the plasma
analogy\cite{Chakrabortybook}. 
On occasions\cite{Girvin}, however, one can find an analytic 
approximation for $g(r)$ based on the fact that it must fulfill generic
sum rules. In addition to these sum
rules and in order to get the desired accuracy, one must provide 
microscopic information specific to the quantum liquid state. This 
information is essential, but, as shown below,
can be obtained from simple considerations which do not necessarily
involve the wave function. Here we find  
$g(r)$ and calculate the energy of spin-polarized and 
spin-unpolarized quantum liquid phases in
the range of filling factors between $\nu=2/3$ and $\nu=1$. 
Exact diagonalizations for $\nu=2/3$, where we certainly expect liquid
phases, and $\nu=4/5$ allow us to confirm the reliability of 
this procedure. Figure \ref{liquid} shows the energy per particle  
of both liquid phases compared to that of the 
solid phase proposed by Brey et al.\cite{Brey}
in a mean-field approximation. Our results confirm the existence of  
liquid phase ground states at $\nu=2/3$ and also at $\nu=4/5$\cite{note1}. 
Most importantly, our results suggest that,
even without considering partial polarization of the liquid phase,
the quantum melting phase transition takes place very close to
$\nu=1$. In fact, it seems to occur near the transition point that
separates the square Skyrme lattice from the triangular 
one\cite{Brey,crystals}.

Following early work by Girvin\cite{Girvin}, we start by noting that the pair
distribution function of a liquid state in the lowest Landau level (LLL)
 can be expanded in terms of 
wave functions for relative states of two particles. Girvin noticed that
it is convenient to recast this expansion in terms of  the pair
distribution function for the $\nu=1$ liquid plus a perturbative series. 
When the liquid is fully polarized 
\begin{equation}
g(r)=1-e^{-r^2/2}+e^{-r^2/4}\sum_{m=1}^\infty \,^{\prime} 
c_m\frac{2}{m!}\left(\frac{r^2}{4}\right)^m,
\label{gp}
\end{equation}
where the prime denotes a sum restricted to odd values of $m$.
The generalized $g(r)$ for fully spin-unpolarized states (i.e., with equal
density of spin-up and spin-down electrons)
\begin{equation}
g(r)= \frac{1}{2}[g_{\uparrow\uparrow}(r)+ g_{\uparrow\downarrow}(r)],
\label{gt}
\end{equation}
is divided into a contribution coming from equal spin 
electrons, $g_{\uparrow\uparrow}(r) [\equiv g_{\downarrow\downarrow}(r)]$, 
and a contribution from opposite spin electrons, 
$g_{\uparrow\downarrow}(r) [\equiv g_{\downarrow\uparrow}(r)]$. It is natural 
to expand $g_{\uparrow\uparrow}(r)$ in a way similar to that in (\ref{gp}). 
It is also convenient to make a similar expansion for  
$g_{\uparrow\downarrow}(r)$, except for the fact that the
sum runs now over all possible non-negative integer values of $m$:
\begin{equation}
g_{\uparrow\downarrow}(r)=1-e^{-r^2/2}+e^{-r^2/4}\sum_{m=0}^\infty 
\tilde c_m \frac{2}{m!}\left(\frac{r^2}{4}\right)^m.
\label{gu}
\end{equation}
If, for all $m$, $c_m \rightarrow 0$ and $\tilde c_m \rightarrow 0$, 
one recovers the partial distribution functions $g_{\uparrow\uparrow}(r)/2$ and 
$g_{\uparrow\downarrow}(r)/2$ of the $\nu=1$ liquid 
ferromagnet with minimum polarization in the direction of the field ($S_z=0$).
Note that, in general, not any set of coefficients 
$c_m$ and $\tilde{c}_m$ represents a good liquid. First of all, at any filling
factor, $c_m \rightarrow 0$ and $\tilde{c}_m \rightarrow 0$ for 
$m\rightarrow \infty$ since $g(r)\rightarrow 1$ for $r
\rightarrow \infty$. Secondly, the pair distribution function must
satisfy certain sum rules which establish relations between the coefficients.
Finally, these coefficients must be consistent with the  
microscopic information one must provide. 

{\em Spin-polarized liquids.} When one applies the charge neutrality, 
perfect screening, 
and compressibility sum rules\cite{Girvin} to the expression (\ref{gp}) 
one obtains the following set of equations:
\begin{eqnarray}
&&\sum_{m=1}^\infty\, ^{\prime} c_m=(1-\nu^{-1})/4 \label{sumrule1} \\
&&\sum_{m=1}^\infty\, ^{\prime} (m+1)c_m=(1-\nu^{-1})/8 \label{sumrule2} \\
&&\sum_{m=1}^\infty\, ^{\prime} (m+1)(m+2)c_m=(1-\nu^{-1})^2/8
 \label{sumrule3} .
\end{eqnarray}
The charge neutrality and perfect screening 
sum rules can actually be derived exactly from general
considerations applicable to any isotropic liquid state in the LLL\cite{sma}. 
In short, they come about due to the fact 
that the number of particles and the total angular momentum are good quantum
numbers for the Hamiltonian. The compressibility sum rule can only be invoked 
as it is if there is a classical plasma analog similar to that for the 
Laughling states\cite{Chakrabortybook}. For the
filling factors considered below this analog exists\cite{ahm}.

Now, in order to incorporate the 
microscopic information related to the short-distance behavior 
into the pair distribution function, we first
express it in a more orthodox way\cite{sma}:
\begin{equation}
g(r)=\frac{1}{\nu^2}e^{-r^2/2}\sum_{m\ne 0}^\infty \langle n_m n_0 \rangle
\frac{r^{2m}}{2^m m!}.
\label{ortho} 
\end{equation}
A series expansion of (\ref{ortho}) allows us to relate the coefficients 
$c_m$ to the correlation factors $\langle n_m n_0 \rangle \equiv 
\langle c^\dagger_mc_m c^\dagger_0c_0 \rangle$ which we try to 
calculate in what follows. 
The particle-hole symmetry dual filling factors of $\nu=1/M$, 
$\nu=1-1/M$,  lie in the range we want to explore here and
we assume a Laughlin state of holes at these filling factors. 
There are no simple wave functions representing these liquids, 
but the same particle-hole symmetry allows us to write
\begin{equation}
\langle n_m n_0 \rangle = 1- \langle h^\dagger_mh_m \rangle - \langle h^\dagger_0h_0
\rangle + \langle h^\dagger_mh_m h^\dagger_0h_0 \rangle
\end{equation}
where the $h^\dagger_m$ and $h_m$ represent creation and annihilation 
operators of holes. For the proposed Laughlin states
\begin{equation}
\langle h^\dagger_mh_m h^\dagger_0h_0 \rangle = 0, \;\;\;\; m\leq M-1 \label{h1}
\end{equation}
which gives $\langle n_m n_0 \rangle=2\nu-1$ for $m\leq M-1$. 
With this result we find $c_m= -(1-\nu^{-1})^2$ for odd $m\leq M-2$ and,
with Eqs. (\ref{sumrule1}), (\ref{sumrule2}), and (\ref{sumrule3}), we
calculate the next three coefficients, setting the rest to zero. 
In Table \ref{table} we present the energy calculated from the
analytic $g(r)$'s so obtained for $\nu=2/3$ and $\nu=4/5$
along with that obtained from exact numerical diagonalizations 
on the spherical geometry for the same filling factors. 
The perfect agreement between these two values confirms the existence of a 
Laughlin liquid state of holes at those filling factors and, moreover, confirms 
the reliability of the analytic pair distribution functions. We calculate now
the energy of Laughlin liquids at any filling factor 
beyond $\nu=4/5$ within the set $\nu=1-1/M$ (stars in Fig. \ref{liquid}).
As $\nu \rightarrow 1$, numerics cannot be employed as a test of accuracy. 
However, all the coefficients must vanish in this limit 
and we expect the expansion (\ref{gp}) to get even more reliable. 
In fact, our results are consistent with the known behavior $\epsilon(\nu
=1)=d\epsilon(\nu)/d\nu|_{\nu \rightarrow 1^-}=-
\frac{1}{2}\sqrt{\pi/2}$\cite{Chakrabortybook} and with the
expected fact that the triangular Skyrme crystal (the two closest 
points to $\nu=1$ in the solid phase) has lower energy than the Laughlin-like
liquid states near $\nu=1$. Figure \ref{liquid} also shows
that the polarized liquid phase beats the solid phase at $\nu=2/3$, but not at 
$\nu=4/5$, in agreement with Ref. \onlinecite{Brey}. 

{\em Spin-unpolarized liquids}. The two first 
previously mentioned sum rules can also be 
invoked for the pair distribution function (\ref{gt}) of a two-component 
liquid. We obtain three sets of equations: 
\begin{eqnarray}
&&\sum_{m=0}^\infty c_m=(1-2\nu^{-1})/4,\;\; 
\sum_{m=0}^\infty \tilde{c}_m= 1/4, \label{sumrule4} \\
&&\sum_{m=0}^\infty (m+1) (c_m+\tilde c_m) =(1-\nu^{-1})/4, \label{sumrule6} 
\end{eqnarray}
where we implicitly consider $c_m=0$ for $m=0,2,4,6,\dots$ 
The first two equations
refer to $g_{\uparrow\uparrow}$ and $g_{\uparrow\downarrow}$, separately. They 
result again from generic considerations\cite{sma} which take into account 
the fact that the number of particles for
each component(spin orientation) is a good quantum number. These two can
be combined into a single one since the energy of the unpolarized phase
depends only on $g_{\uparrow\uparrow}+g_{\uparrow\downarrow}$, i.e., on
$c_m+ \tilde c_m$. The compressibility sum rule has not been invoked this time
for we are not aware of any two-especies plasma analog for the filling 
factors considered above. Its existence cannot be discarded though.
 
Next, we provide the microscopic details of the
unpolarized liquid states we propose. An 
expansion like (\ref{ortho}) including the spin degree
of freedom allows us again to relate the coefficients $c_m$ and $\tilde c_m$ to
the correlation factors $\langle n_{m\uparrow} n_{0\uparrow}\rangle (\equiv
\langle n_{m\downarrow} n_{0\downarrow}\rangle)$ and 
$\langle n_{m\uparrow} n_{0\downarrow}\rangle (\equiv
\langle n_{m\downarrow} n_{0\uparrow}\rangle)$. 
The duality between $\nu$ and $1-\nu$ in previous section allowed
us to propose sensible liquid ground state wave functions 
at filling factors $\nu=1-1/M$,
but, here, the spin degree of freedom breaks such a duality. 
However, note that one can always find an appropriate hard-core interaction 
model for which the polarized ground state satisfies Eq. (\ref{h1}), 
whether or not its  wave function is explicitly known. 
Similarly, the hard-core interaction model $V_0 \gg V_1>V_2=V_3\dots=0$ 
guarantees that the unpolarized ground state
satisfies $\langle n_{0\uparrow} n_{0\downarrow}\rangle = 
\langle n_{0\downarrow} n_{0\uparrow}\rangle=0$ for $\nu\leq 1$. It also
guarantees the upper limit  
\begin{equation}
\langle n_1 n_0 \rangle \equiv 2\langle n_{1\uparrow} n_{0\uparrow}\rangle+
2\langle n_{1\uparrow} n_{0\downarrow}\rangle  \leq (2\nu-1),
\label{coefu}
\end{equation}
which, as follows from previous arguments, is the value associated 
to the fully polarized state at filling factors $2/3 \leq \nu\leq 1$ 
for any $S_z$.
The inequality reflects the freedom for a pair of particles to occupy
states with even relative angular momenta. We propose the following
factorization: $\langle n_1 n_0 \rangle = f(\nu)(2\nu-1)$ with $ f(\nu)\leq 1$. 
In general, $f(\nu)=\alpha + (1-\alpha) \nu$ since we do not expect terms
proportional to $\nu^3$ or higher to be relevant in a density-density 
correlation function and the condition  $f(\nu=1)=1$ must be fulfilled.
Thus we set $\tilde c_0=0$ and $c_1+\tilde
c_1=2(1-\alpha)(1-\nu^{-1})-2\alpha(1-\nu^{-1})^2$ and, 
using Eqs. (\ref{sumrule4}) and (\ref{sumrule6}), we can calculate $\tilde
c_2$ and $c_3+\tilde c_3$. We especifically set $\alpha=25/8-2\sqrt{2}$ 
which builds a $g(r)$ from which we
reproduce the expected behavior of the energy for a very dilute state of 
skyrmions at vanishing Zeeman coupling,
$d\epsilon(\nu)/d\nu|_{\nu\rightarrow 1^-} = -\frac{1}{4}\sqrt{\pi/2}$.
Moreover, this choice of $\alpha$
is also consistent with our best numerical estimate of 
$\langle n_1 n_0 \rangle$ for $\nu=2/3$. As Table \ref{table} shows, the energy 
obtained from the analytical expression of 
$g(r)$ for $\nu=2/3$ and $\nu=4/5$ and the exact results  
for the real Coulomb interaction are remarkably close. The solid phase energy
(see Fig. \ref{liquid}) lies clearly above the exact and estimated values, which
proves the existence of unpolarized liquid phases at these
filling factors. (Note that the energy of the polarized phases are always 
above the corresponding ones of the unpolarized phases, but a
{\em partially} polarized liquid at $\nu=4/5$ is possible and might have
lower energy for the value of the Zeeman coupling considered). One can 
interpret these unpolarized states as $\nu=1/3$ and $\nu=1/5$ hole liquid 
states, with each hole being ``accompanied'' by one and two spin flips 
(i.e., spin waves), respectively. 
Although we cannot present a rigorous argument, it is tempting to interpret 
this hole-spin waves association as a skyrmion\cite{Jacob}, and call 
these states Skyrme liquids. 

In order to address the quantum melting phase transition we need
the energy at any filling factor $2/3\leq\nu\leq 1$:
\begin{equation}
\epsilon=-\frac{\nu}{2}\sqrt{\frac{\pi}{2}}\left[1-\frac{1}{2}(1-\nu^{-1})+
(\frac{25\sqrt{2}}{64}-\frac{1}{2}) (1-\nu^{-1})^2\right].
\label{eu}
\end{equation}
Although, as required, this expression reproduces known results fairly well, 
we have to stress that {\em its derivative is continuous as a function of $\nu$}. 
In other words, it does not capture the expected gaps in the chemical potential 
at the main filling fractions. Experimentally, however,
only the gap at $\nu=2/3$ 
is clearly visible for fully-polarized phases. Towards $\nu=1$
the gaps at $\nu=5/7$ and $\nu=4/5$ are the next in importance, but they
barely show up in the longitudinal or Hall resistivities
even at the lowest temperatures. In addition, 
the gaps are smaller in unpolarized phases\cite{Chakrabortybook} so we expect
the analytical expression (\ref{eu}) to capture the overall behavior
of the real $\epsilon(\nu)$ of a spin-unpolarized liquid, 
particularly close to $\nu=1$.
Before drawing conclusions about the quantum melting phase transition
from Fig. \ref{liquid} one should keep in mind that quantum
fluctuations have not been considered in the solid phase calculation
and that they are bound to lower the energy of this phase.
The magnitude of this correction has been carefully estimated for the Wigner 
crystal\cite{Yi} and it is not unreasonable to expect an energy shift 
of the same order of magnitude for the Skyrme crystal. Borrowing the results
from Ref. \onlinecite{Yi} we estimate the shift to be $-0.001 e^2/\ell$ 
for $\nu=4/5$, decreasing in absolute value as $\nu\rightarrow 1$. 
In our favor we must also say that 
partially polarized liquid phases, which have not been considered here,
are expected to lower the ground state energy for typical values of
the Zeeman energy. All considered, our results seem 
to place the quantum melting phase transition very close to $\nu=1$. As
expected, the triangular Skyrme crystal
dominates sufficiently close to $\nu=1$, but the transition point to the square
lattice ($\nu\approx 0.96$) lies near the one  
where the unpolarized liquid phase begins to take over ($\nu\approx 0.92$).  

There are still many open
questions regarding the nature of the zero-temperature melting phase
transition over which this work does not shed any light. Disorder, 
for instance, plays a major role in this matter and has not been considered.  
Still, our results cast a caveat in what refers to the range 
of existence of the square-lattice Skyrme crystal and, in addition, provide
an alternative description for the electronic ground states
in the range of filling factors $2/3\leq\nu\leq 1$.

We would like to thank Luis Brey for providing us with the data for the 
solid phase. We also acknowledge fruitful discussions with C. Tejedor, P.
Tarazona, E. Chac\'on, J. Fernandez-Rossier, and L. Mart\'{\i}n-Moreno. 
This work has been funded by MEC of Spain under contract No. PB96-0085.


\begin{table}
\begin{center}
\begin{tabular}{|c|cccccc|c|}
$\nu$&$N=4$&$N=6$&$N=8$&$N=10$&$N=12$&$N\rightarrow\infty$&$g(r)$ \\ \hline
2/3 (P)&-0.5501&-0.5396&-0.5341&-0.5310&-0.5289&-0.518&-0.518\\
2/3 (U)&-0.6010&-0.5744&-0.5618&-0.5545&       &-0.527&-0.528\\ \hline
4/5 (P)&-0.5804&       &-0.5649&       &-0.5607&-0.552&-0.552\\
4/5 (U)&-0.6580&       &-0.6069&       &-0.5923&-0.56(7)&-0.566    
\end{tabular}
\end{center}
\caption[]{Energy per particle (in units of $e^2/\ell$)
obtained from exact diagonalizations on
a sphere for finite number of particles, $N$, at filling factors
$\nu=2/3$ and $\nu=4/5$ (when possible, values for $N>12$ have been
obtained, but are not shown here). 
The spin-polarized and spin-unpolarized calculations
are denoted by P and U, respectively. We also include the  thermodynamic 
value obtained from a $1/N$ cuadratic extrapolation
and the energy obtained from the pair 
distribution functions described in the text. 
The last digit of the extrapolated value for $\nu=4/5$
(U) cannot be taken at face value due to the small number of points available
for the extrapolation.}
\label{table}
\end{table}                     


\end{document}